\documentstyle[12pt,psfig]{article}

\newcommand{\epsfig}{\psfig}
\newcommand{\bh}{{\bar h}}
\newcommand{\rg}{\sqrt{g}}

\newcommand{\p}{\partial}

\newcommand{\VRie}{{V}^{\mu\nu}_{(R_{\mu\rho\nu\sigma})^2}}
\newcommand{\VRic}{{V}^{\mu\nu}_{(R_{\mu\nu})^2}}
\newcommand{\VRs}{{V}^{\mu\nu}_{(R^2)}}

\newcommand{\VR}{{V}_R}

\newcommand{\val}[2]{{\left.{#1}\right|_{#2}}}
\newcommand{\bea}{\begin{eqnarray}}
\newcommand{\eea}{\end{eqnarray}}
\newcommand{\be}{\begin{equation}}
\newcommand{\ee}{\end{equation}}
\newcommand{\eps}{\epsilon}
\newcommand{\la}{\langle}
\newcommand{\ra}{\rangle}
\newcommand{\Gam}{\Gamma}
\newcommand{\vart}{\vartheta}
\newcommand{\del}{\delta}
\newcommand{\Rie}{R_{\mu\rho\nu}{}^{\sigma}R^{\mu\rho\nu}{}_{\sigma}}
\newcommand{\C}{C_{\mu\rho\nu}{}^{\sigma}C^{\mu\rho\nu}{}_{\sigma}}
\newcommand{\Ric}{R_{\mu\nu}R^{\mu\nu}}
\newcommand{\Rs}{R^2}

\begin{document}


\begin{titlepage}
\thispagestyle{empty}
\vspace*{-0.568in}
\begin{flushleft}
UPR-664T\\
LMU-TPW  96-14
\end{flushleft}
\begin{center}
\bigskip\bigskip
\vspace*{2cm}
{\LARGE{ \bf Higher-Derivative Gravity in \\String Theory}}
\vskip2cm
Kristin F\"orger$^1$, Burt A. Ovrut$^2$, Stefan J. Theisen$^1$, Daniel 
Waldram$^2$
\vskip2cm
{\sl $^1$Sektion Physik, Universit\"at M\"unchen \\
Theresienstr. 37, 80333 M\"unchen, Germany}

\vspace{0.8cm}
{\sl $^2$Department of Physics, University of Pennsylvania \\
Philadelphia, PA 19104-6396, USA}
\end{center}
\vskip1in


\begin{abstract}
We explicitly extract the structure of higher-derivative curvature-squared 
terms at genus $0$ and $1$
in the $d=4$ heterotic string effective action compactified on
symmetric orbifolds by computing on-shell S-matrix superstring amplitudes.
In particular, this is done within the context of calculating the graviton 
$4$-point amplitude. We also discuss the moduli-dependent 
gravitational threshold corrections to 
the coupling associated with the $CP$ even quadratic curvature terms.  
\end{abstract}
\mbox{}
\end{titlepage}
\newpage
\setcounter{page}{1}


\subsection*{1. Introduction}

It is a well known fact that perturbative string theory
contains the pure Einstein-Hilbert action. A priori, there is no reason why
higher-derivative gravitational terms could not also be present in the action
of strings and superstrings. Recently, it has been shown that higher-
derivative supergravity terms can provide a new mechanism for supersymmetry
breaking in $d=2$ and $d=4$ supergravity models \cite{abd} and, hence, 
such terms could be fundamentally important. An attempt to determine this
question was made within the context of superstring compactifications on
${\bf Z}_N$ symmetric orbifolds in \cite{clo}. It was demonstrated, using
light field one-loop radiative corrections to the string effective Lagrangian,
that higher-derivative supergravitational terms can arise in string theory
and, importantly, that they need not be in the topological Gau{\ss}-Bonnet
combination. However, this discussion was not conclusive since the radiative
contributions of the infinite tower of massive states was not included in the
calculation. It is possible, although it was argued that it was unlikely, that
these massive contributions would cancel the higher-derivative terms generated
by the light fields. It seems clear that the only way to resolve this issue is
to do complete genus-0 and genus-1 superstring amplitude calculations, since 
such
calculations include both massless and massive states. One would then attempt
to construct the string action associated with these amplitudes. In this way
one could explore
the interesting question of whether or not string theory can completely 
fix the curvature squared terms in its action, at least  
to the genus-1 level. The answer to this problem is the subject of this paper.

\subsection*{2.  String $S$-Matrix}

The perturbative $S$-matrix approach consists of finding a local 
gauge-invariant effective Lagrangian ${\cal L}$ whose associated
$S$-matrix elements  coincide with the string $S$-matrix. More precisely this 
means that the effective action, which is
a power series in the external momenta $k^2 \alpha'$ and the string coupling 
constant $g_s$, reproduces
the kinematic structure of the string amplitudes.
 
We start by giving the most general ansatz for the bosonic part of the  
$CP$ even gravitational Lagrangian
\be\label{eq:ea}
{\cal L}=\rg\Big\{\frac{1}{2 \kappa^2} R+a\; \Rie+b\; \Ric+c\; \Rs\Big\}
\ee
where $\kappa$ is the four dimensional 
gravitational coupling constant. In the following we set $2\kappa^2=1$ and
reintroduce it later for special purposes. We are not considering the  $CP$-odd 
part of the gravitational action here,
which is given by a topological invariant,  the first Pontrjagin class in four 
dimensions. 
The aim is to try to fix the unknown coefficients $a$, $b$ and $c$ by 
calculating string amplitudes. 

In order to find the appropriate $n$-point string amplitudes, we expand 
the above Lagrangian
around the flat background metric $g_{\mu\nu}=\eta_{\mu\nu}+h_{\mu\nu}$. 
The inverse metric is then given by a power series
expansion $g^{\mu\nu}=\eta^{\mu\nu}-h^{\mu\nu}+(h^2)^{\mu\nu}-\ldots$ and 
$h^{\mu\nu}=\eta^{\mu\rho}\eta^{\nu\sigma}h_{\rho\sigma}$.
Furthermore, in order to correspond to the on-shell fluctuations described by
the string graviton-vertex operator, we demand that
$h_{\mu\nu}$ satisfies the harmonic gauge conditions for
a massless spin $2$ field, namely  $\Box h_{\mu\nu}=0$, 
$\p^{\mu}h_{\mu\nu}=0$ and the tracelessness  condition $h=0$. 
Expanding the curvature terms in equation (\ref{eq:ea}) in terms of $h$,
the first non-vanishing contributions 
arise at the $3$-point level \cite{grs}. These are
\bea\label{Rh3}
\val{\rg R}{h^3}&=&h_{\mu\nu}\, (h,{}^{\mu\nu} h)+2 h_{\mu\nu},
{}^{\sigma} \;h^{\nu\rho},{}^{\mu}  
\;h_{\rho\sigma}\nonumber\\
&\to& (k_2 \eps^1 k_2) (\eps^2\eps^3)+2 
(k_3\eps^2\eps^3\eps^1k_2)+\mbox{cyclic perm.}\\
\val{\rg\Rie}{h^3}&=&h_{\mu\nu},_{\rho}{}^{\lambda}\, h_{\rho\sigma} 
\,h^{\sigma\lambda},^{\mu\nu}\nonumber\\
&\to& (k_1\eps^3k_1)(k_3\eps^2\eps^1 k_3)
+\mbox{cyclic perm.}
\eea
where the notation 
$(h,{}^{\mu\nu}\;h)=(\p^{\mu}\p^{\nu}h_{\rho\sigma})\,h^{\rho\sigma}$ has been 
used.
We have also introduced the tranformation to the momentum space, replacing 
$i\p \to k$, and replaced 
$h_{\mu\nu} \to\eps_{\mu\nu}$, where $\eps_{\mu\nu}$ 
is the graviton polarization tensor. Note that this polarization tensor
satisfies the same differential and tracelessness conditions as $h_{\mu\nu}$. 
Unfortunately, the kinematic structure of $\Ric$ and $\Rs$ are such that they 
must vanish when expanded in 
$h_{\mu\nu}$ to the $3$-point level. That is
\begin{equation}
\val{\rg\Ric}{h^3}=\val{\rg\Rs}{h^3}=0
\label{eq:burt}
\end{equation}
Therefore, a $3$-point string amplitude with 
three on-shell external gravitational
vertex operators can  produce only the first two terms of the effective 
Lagrangian 
${\cal L}$. The other two terms may exist in the effective Lagrangian, but this
cannot be determined at the $3$-point level.
Fortunately, at the level of $4$-point amplitudes (and higher), the Ricci 
squared and curvature scalar squared terms no longer vanish.
That is, at the level of $4$-point amplitudes
$\val{\rg\Ric}{h^4}$ and $\val{\rg \Rs}{h^4}$ are non-zero.
In the expansion up to order $h^4$ of the curvature squared terms, it is 
sufficient for our purposes to isolate only certain terms, which are 
particular combinations of the polarization tensors, namely 
\bea\label{Rie}
\val{\rg\Rie}{h^4}&=&\Big(\frac{1}{8} u^2+\frac{3}{2} s^2 
\Big)E_2+\Big(\frac{1}{8} t^2-\frac{1}{4} s^2\Big) E_1 + \cdots \\
\label{Ric}\val{\rg\Ric}{h^4}&=&\frac{1}{4} s^2 E_2+\Big(\frac{1}{16} 
t^2+\frac{1}{8} s^2\Big) E_1 + \cdots\\
\label{Rs}\val{\rg\Rs}{h^4}&=&\frac{9}{16} s^2 E_1 + \cdots
\eea
where $s=-2 k_1\cdot k_2$, $t=-2 k_1\cdot k_3$ and $u=-2 k_1\cdot k_4$ are the 
Mandelstam 
variables and we introduce
 $E_1=(\eps_1\,\eps_2)(\eps_3\eps_4)$ and
$E_2=(\eps_1\eps_2\eps_4\eps_3)$, using the matrix notation 
$(\eps_1\eps_2)=\eps_1{}_{\mu\nu}\eps_2{}^{\mu\nu}$.

At tree level, both the $3$- and $4$-point  graviton amplitudes for the 
heterotic 
string in $d=4$ have already been studied
 in e.g.  \cite{grs},\cite{ltz},\cite{dr},\cite{cn}.
The ${\cal O}(k^2)$ part of the $3$-point tree level amplitude 
corresponds to the curvature scalar $R$ and, therefore, reproduces
the Einstein-Hilbert action. Additionally, it gives the relation 
$2\kappa^2=g_s^2 
\alpha'$ beween the gravitational and string coupling constants. The
${\cal O}(k^{4})$ terms in the $3$-point amplitudes can only give rise 
to one of the curvature squared terms, as discussed above, so we 
won't discuss them here.
All three curvature squared terms arise at order ${\cal O}(k^4)$ in 
the four-graviton amplitude. If we again 
restrict ourselves to only the terms involving the particular polarization 
combinations $E_1$ and $E_2$, the ${\cal O}(k^4)$ contribution of 
the $4$-point amplitude is found to be
\be
{\cal A}_{4g}^{tree}=3 g_s^2\Big\{\Big(-\frac{1}{2}s^2+t^2\Big)E_1+u^2 
E_2\Big\} \equiv 3 g_s^2 K_{4g}^{tree}
\ee
Comparing this amplitude with the effective Lagrangian ${\cal L}$, one realizes 
that there is no combination of
curvature squared terms (\ref{Rie}),(\ref{Ric}) and (\ref{Rs})
that reproduces the kinematic structure of ${\cal A}_{4g}^{tree}$. This fact 
is not surprising since string amplitudes
include $1PR$ exchange graphs with massless poles  
as well as $1PI$ graphs. It follows that one has to perform the appropriate 
field theory subtractions in order to relate the string result to the
effective action. This will be done for 
the tree and the 1-loop amplitudes in the next section.

Proceeding in a similar fashion to the genus-zero case, we now go to the 
one-loop level and calculate both the $3$- and $4$-point graviton amplitudes 
on a world-sheet torus for the heterotic string in $d=4$ with a given vacuum.
The general expression for the $CP$ even $n$-point amplitude is \cite{min,bk}
\be
{\cal A}_n^{even}=g_s^n\sum_{(s_1,s_2)\atop even} (-)^{s_1+s_2}\, 
\int_{\tau\in\Gam}\frac{d^2 \tau}{\tau_2}\,Z(\tau,\bar{\tau},{\bf s})
\int\prod_{i=1}^n d^2  z_i \la \prod_{i=1}^n V^{(0)}(z_i,\bar{z}_i)\ra_{\bf s}
\ee
where ${\bf s}=(s_1,s_2)$ characterizes the spin structures, which take the 
values $0$ 
and $1$ for the $NS$ and $R$ sectors, respectively.
The integration region for the modulus of the torus $\tau=\tau_1+i\tau_2$ is 
the fundamental region $\Gam=\{\tau\ |\ |\tau_1|\le\frac{1}{2},|\tau|\ge 1\}$.
The factor $(-)^{s_1+s_2}$ comes from the fermionic partition function 
$Z_{\psi}$ in the light 
cone gauge by demanding that the sum over the spin structures of $Z_{\psi}$ is 
modular invariant.
$Z(\tau,\bar{\tau},{\bf s})$ is the partition function in light cone gauge for 
the heterotic string, given by
\be
Z(\tau,\bar{\tau},{\bf s})={\rm Tr}\Big((-)^{s_2 F} 
q^{H-\frac{1}{2}}\bar{q}^{\bar{H}-1}\Big)=Z_{\psi} Z_X Z_{X_0} Z_{int}
\ee
where $F$ is the fermion number and $q=e^{2\pi i\tau}$.
 $Z_{X_0}=\frac{1}{2(2\pi)^4\tau_2^2}$ is the contribution
from the bosonic zero modes, $Z_X=\frac{1}{|\eta(\tau)|^4}$ is the bosonic 
partition function where the Dedekind $\eta$ function is defined as 
$\eta(\tau)=q^{1/24}\prod_{n=1}^{\infty} (1-q^n)$, and the partition function 
for one complex fermion in the light cone gauge  is 
$Z_{\psi}=\frac{\vart_{\alpha}(0,\tau)}{\eta(\tau)}$ where $\vart_{\alpha}$ 
are the Riemann theta functions for  $\alpha=2,3,4$ corresponding to
the spin structures $(s_1,s_2)=(1,0),(0,0),(0,1)$ respectively.
The term $Z_{int}$ is the partition function for the six-dimensional internal 
compact manifold.
The super-ghost charges in the $CP$ even part of the string amplitude have to 
add up to 
zero. Therefore, the vertex operators are taken to be in the zero ghost picture 
$V^{(0)}$. In this picture, the graviton vertex operator looks like: 
\be
V_g^{(0)}(z,\bar{z})=:\eps_{\mu\nu}\,\bar{\p}X^{\mu}\Big(\p X^{\nu}+i k\cdot 
\psi \psi^{\nu}\Big) \,e^{i k\cdot X}:
\ee
Here $X^{\mu}$ are the free world-sheet bosons and $\psi^{\mu}$ the left-moving 
supersymmetric partners.

Let us start by calculating the ${\cal O}(k^2)$ piece of the $3$-point 
graviton amplitude, which is associated to the $R$ term.
We expand the three-point correlation function of three-graviton vertex 
operators in the zero ghost picture using Wick
contractions.  By supersymmetry, terms that are independent of the spin 
structure give zero after summing over spin structures.
Contractions of the exponential $e^{i k\cdot X}$ give 
$\la :\prod_j e^{i k_j\cdot X_j}:\ra=\prod_{i<j}|\chi_{ij}|^{1/2\; k_i\cdot 
k_j}$  
where we have set $\alpha'=\frac{1}{2}$ and 
\be
\chi_{ij}\equiv\chi(z_{ij},\tau)=2\pi e^{-\pi ({\rm Im}z_{ij})^2/{\rm 
Im}\tau_2}\left|\frac{\vart_1(z_{ij},\tau)}{\vart_1'(0,\tau)}\right|
\ee
which is related to the bosonic Green function on the torus by 
$G^B_{ij}=-\frac{1}{4} \ln|\chi_{ij}|^2$.
To get the whole kinematic structure of $\rg R$ as in expression
(\ref{Rh3}), one also has to take into account contributions
coming from the ${\cal O}(k^4)$ terms by a `pinched off' integration 
\cite{min,ms,bk}. 
These terms arise in the limiting case $z_{ij}\to 0$, for which we get
\be
|\chi_{ij}|^2\to |z_{ij}|^2\,, \hspace{1cm}\bar{\p}G_{ij}^B\to 
-\frac{1}{4\bar{z}_{ij}}\,,\hspace{1cm}
\p G_{ij}^B\to -\frac{1}{4 z_{ij}}
\ee
The world-sheet integral over the region $|z_{ij}|<\eps$ then yields a pole in 
$k_i\cdot k_j$, after analytic continuation:
\be
\int_{|z_{ij}|<\eps} d^2z_{ij} \frac{|z_{ij}|^{1/2\; k_i\cdot k_j}}{16 
|z_{ij}|^2} 
\simeq \frac{4\pi}{16\,k_i\cdot k_j}.
\ee
where we assumed that $\frac{1}{2} \left|k_i\cdot k_j\right| 
\ll\left|\frac{1}{\ln\eps}\right|$.
After performing one world-sheet integration we get the following 
${\cal O}(k^2)$ expression: 
\bea
\lefteqn{\int \prod_{i=1}^3 d^2z_i \val{\la V_g^{(0)}(z_i,\bar{z}_i)\ra}
{{\cal O}(k^2)}\simeq}\nonumber\\ 
& & \int d^2z_1\int d^2z_2 (G_{23}^F)^2 \bar{\p}^2G_{12}^B [(k_2\eps^1k_2)
(\eps^2\eps^3)+2(k_3\eps^2\eps^3\eps^1k_2)] + \mbox{cyclic perm.}
\eea
where $G^F_{ij}=G^F(z_{ij})$ and 
$G^F(z)=\frac{1}{4}\,\frac{\vart_1'(0,\tau)\,\vart_{\alpha}(z,\tau)}
{\vart_{\alpha}(0,\tau)\,\vart_1(z,\tau)}$
is the fermionic Green function on the torus. If we use the fact that $(G^F)^2$ 
can be expressed as 
\be
(G^F(z))^2= \frac{1}{16}\,\left(-\p^2\ln\vart_1(z,\tau)+4\pi 
i\p_{\tau}\ln\eta(\tau)-e_{\alpha}\right)
\ee
where $e_{\alpha}=-4 \pi i \p_{\tau}\ln Z_{\psi}$, we can replace $(G^F)^2$ by 
the spin dependent part $(-\frac{e_{\alpha}}{16})$
because the other two terms of $(G^F)^2$ will yield a zero result for the 
amplitude after summing over even spin structures.
Therefore, there remains only one world-sheet integral to be done. But since 
\be\label{bgint}
\int d^2z \,\bar{\p}G^B=\int d^2z\, \bar{\p}^2G^B=0
\ee
we find that the one-loop coupling in front of $\rg R$ vanishes. This means 
that there 
is no renormalization of Newton's constant for heterotic strings
in $d=4$, a fact  which was also noted in \cite{agn1} and \cite{kiko}, 
whereas for  type II strings in $d=4$ it does get renormalized 
at one-loop string level, because the fermionic zero modes can be saturated by 
contracting left- and right-moving 
fermions \cite{kiko}.

We now turn to the curvature squared terms and calculate the 
${\cal O}(k^4)$ terms of the one-loop $4$-graviton amplitude for 
a $d=4$ heterotic string. By doing so, we hope to determine whether or not 
the coefficients of the curvature squared terms of the one-loop 
effective action can be uniquely  fixed.
The four graviton vertex correlation function is:
\be
\la \prod_{i=1}^4 V_g^{(0)}(z_i,\bar{z}_i)\ra=\prod_{i<j}|\chi_{ij}|^{1/2\; 
k_i\cdot k_j}\,\big(X(k,\eps,z)+Y(k,\eps,z)+Z(k,\eps,z)\big)
\ee
where $X(k,\eps,z)$, $Y(k,\eps,z)$ and $Z(k,\eps,z)$ are polynomials of $k$, 
$\eps$ and bosonic- and fermionic
Green functions, including a $4$, $6$ and $8$ fermion 
correlation function, respectively.

If we concentrate on ${\cal O}(k^4)$ contributions, we also have 
to expand the contractions of the exponential
\be
\prod_{i<j}|\chi_{ij}|^{1/2\; k_i\cdot k_j}=1-\sum_{i<j} k_i\cdot k_j 
G_{ij}^B+\frac{1}{2}\big(\sum_{i<j} k_i\cdot k_j G_{ij}^B\big)^2-\ldots
\ee
 which is valid only if $|z_{ij}|>\eps$ because of the singularity of $G^B$ at 
the origin. For $z_{ij}\to 0$ one has to use the 
pinched off integration. 

Performing the world-sheet integrals one realizes that many of the integrals 
give a zero result because of (\ref{bgint}). Further, 
integrals such as $\int d^2z_1 \bar{\p}^2_1
G_{12}^B \,cs(2 K z_{12}) \,cs(2K z_{13})$ also vanish as a result of the 
(quasi) periodicity of 
the Jacobi elliptic function  $cs(2Kz)$, where 
$K=\frac{\pi}{2}\vart_3^2(0,\tau)$ and 
 which is  defined as 
$cs(2 K 
z)=\frac{\vart_2(z,\tau)\vart_4(0,\tau)}{\vart_1(z,\tau)\vart_3(0,\tau)}$. So 
$cs(2 Kz)$
is related to the fermionic Green function for $\alpha=2$. Finally, we find
that the $E_1$ and $E_2$ dependent part of the one-loop
string amplitude is
\be
{\cal A}_{4g}^{even}=-\frac{3}{2}g_s^4 K_{4g}^{1-loop} \,\sum_{{\bf s}\, 
even}(-)^{s_1+s_2}\int\frac{d^2\tau}{\tau_2}Z(\tau,\bar{\tau},{\bf s})\int
\prod_{i=1}^4 d^2z_i\, (G_{34}^F)^2 v^2 (\bar{\p}G_{34}^B)^2
\ee
where $v=\frac{\pi}{4{\tau}_2}$ and $K_{4g}^{1-loop}
=[-s^2 E_1+(2s^2+u^2) E_2]$. 
The remaining world-sheet integral is given by the following result
\be
\int d^2z 
(\bar{\p}G^B(z))^2=-\frac{i\pi\tau_2}{8}\p_{\bar{\tau}}\ln(\tau_2\bar{\eta}^2)=-
\frac{\tau_2}{16}\hat{G}_2(\bar{\tau})
\ee
where we used the heat equation and periodicity of $\vart_1(z,\tau)$, and the 
Eisenstein function of weight $2$,
$\hat{G}_2(\tau)=G_2(\tau)-\frac{\pi}{{\rm Im}\tau}$.
Thus we get for the amplitude
\be
{\cal A}_{4g}^{even}=-\frac{6 i}{\pi}\Big(\frac{g_s}{16}\Big)^4\, 
K_{4g}^{1-loop}\,\sum_{{\bf s}\, 
even}(-)^{s_1+s_2}\int\frac{d^2\tau}{\tau_2}\frac{1}{
|\eta(\tau)|^4}Z_{int}\,\p_{\tau}Z_{\psi}\,\hat{G}_2(\bar{\tau})
\ee
Note that $K_{4g}^{1-loop}\neq K_{4g}^{tree}$. Thus either the one-loop 
corrections to the effective action are not proportional to the tree-level 
effective action, or the one-loop string amplitude contains different field 
theory subtractions. 

The $\tau$ integral can be performed after fixing the internal sector, as, for
example, a symmetric orbifold. This will be done after the field 
theory subtraction.

\subsection*{3. Field Theory Comparison}

In the previous section, we computed 4-point string amplitudes and  
obtained particular 
kinematic structures for the ${\cal O}(k^4)$ parts.
 As we already mentioned, no combination of curvature squared terms that are 
expanded to the order $h^4$ can reproduce either $K_{4g}^{tree}$ or
$K_{4g}^{1-loop}$. The way to proceed is to take the Lagrangian in 
(\ref{eq:ea}), 
calculate all  possible field theory contact and exchange 
graphs 
that contribute to ${\cal O}(k^4)$, and subtract them from the string 
amplitude expression to get the $1PI$ effective action.

To do this we introduce $\bh_{\mu\nu}$ as an off-shell graviton field.
That is, unlike $h_{\mu\nu}$ which is constrained to satisfy the equation of
motion, $\bh_{\mu\nu}$ is an arbitrary fluctuation.
We determine the vertices for two on-shell and one off-shell field. Expanding 
the scalar curvature
$\val{\rg R}{h^2\bh}=\bh_{\mu\nu} {\VR}$ we get the vertex
 \be
{\VR}=\frac{3}{8}\eta^{\mu\nu}(h,_{\rho}\;h,^{\rho})-\frac{1}{4}
(h,^{\mu}\;h,^{\nu})-\frac{1}{2}(h\;h,^{\mu\nu})-\frac{1}{2}h^{\mu\alpha},
_{\rho}\;h^{\nu}{}_{\alpha},^{\rho}+\ldots
\ee
For the moment we are only interested in terms having the form 
$(h,^{\mu}\;h,^{\nu})$, $(h,^{\mu\nu}\;h)$, $h^{\mu\alpha},_{\rho}\;h^{\nu}
{}_{\alpha},^{\rho}$ and $(h,_{\alpha}\;h,^{\alpha})$ because in exchange 
graphs it will be these terms that 
 contribute to those kinematic 
terms depending on $E_1$ and $E_2$, which we picked out for the comparison 
with the $4$-graviton string amplitude.
The vertex for $\Rie$ can be obtained by taking 
$\val{\rg\Rie}{h^2\bh}=\bh_{\mu\nu}{\VRie}$, the vertex for $\Ric$ 
by $\val{\rg\Ric}{h^2\bh}=\bh_{\mu\nu}{\VRic}$ and the $\Rs$ 
vertex by
$\val{\rg\Rs}{h^2\bh}=\bh_{\mu\nu}{\VRs}$. We find 
\bea
{\VRie}&=&\frac{1}{2}\eta^{\mu\nu}(h,^{\rho\sigma}\;h,_{\rho\sigma})
-5(h^{\mu\alpha},_{\rho\sigma}\;h^{\nu}{}_{\alpha},^
{\rho\sigma})-(h,^{\mu\nu}{}_{\rho}\;h,^{\rho})+\ldots\\
{\VRic}&=&(h,^{\mu\rho}\;h,^{\nu}{}_{\rho})+\frac{1}{2}(h,^{\rho}
\;h,^{\mu\nu}{}_{\rho})-(h^{\mu\alpha},_{\rho\sigma}\;h^{\nu}{}_{\alpha},
^{\rho\sigma})+\dots\\
{\VRs}&=&3(h,^{\mu\rho}\;h,^{\nu}{}_{\rho})+3(h,^{\rho}
\;h,^{\mu\nu}{}_{\rho})-3\eta^{\mu\nu}(h,^{\alpha\rho}\;h,_{\alpha\rho})+\ldots
\eea
 
Having defined the vertices, we now introduce the internal propagator for the 
exchange graphs. Expanding (\ref{eq:ea})
to quadratic order and inverting the kernel, one finds the following
propagator
\be
D_{\mu\nu\rho\sigma}=-\left\{\frac{P_{\mu\nu\rho\sigma}^{(2)}}
{k^2(1+k^2(b+4a))}
-\frac{P_{\mu\nu\rho\sigma}^{(0)}}{2k^2(1+2 k^2(-3a-b-3c))}\right\}
\ee
where $P_{\mu\nu\rho\sigma}^{(2)}$ and $P_{\mu\nu\rho\sigma}^{(0)}$ are 
transverse projectors for spin-$2$ and spin-$0$, 
respectively
\bea
P_{\mu\nu\rho\sigma}^{(2)}&=&\frac{1}{2}(\theta_{\mu\rho}\theta_{\nu\sigma}
+\theta_{\mu\sigma}\theta_{\nu\rho})-P_{\mu\nu\rho\sigma}^{(0)}\\
P_{\mu\nu\rho\sigma}^{(0)}&=&\frac{1}{3}\theta_{\mu\nu}\theta_{\rho\sigma}
\eea
where $\theta_{\mu\nu}=\eta_{\mu\nu}-\omega_{\mu\nu}$ and 
$\omega_{\mu\nu}=\frac{\p_{\mu}\p_{\nu}}{\Box}$.
We expand the propagator for small momenta and get 
$D_{\mu\nu\rho\sigma}=\Pi_{\mu\nu\rho\sigma}+\tilde{\Pi}_{\mu\nu\rho\sigma}$,
where the first term is the usual graviton propagator 
\be
\Pi_{\mu\nu\rho\sigma}=-\frac{1}{k^2}\Big(P_{\mu\nu\rho\sigma}^{(2)}
-\frac{1}{2}P_{\mu\nu\rho\sigma}^{(0)}\Big)
\ee
 and the second term is a correction to the graviton propagator:
\be
\tilde{\Pi}_{\mu\nu\rho\sigma}=(b+4a)P_{\mu\nu\rho\sigma}^{(2)}
+(a+b+3c)P_{\mu\nu\rho\sigma}^{(0)}.
\ee

We can now calculate  the field theory  exchange graphs using $\Pi$ and 
$\tilde{\Pi}$ as internal propagators.
The four-graviton tree-level  amplitude is reproduced by a tree-level contact 
graph Fig.1 and tree-level
exchange graphs Fig.2.
\begin{figure}[htb]
\begin{center}
\begin{minipage}{4cm}
\begin{picture}(0,0)%
\epsfig{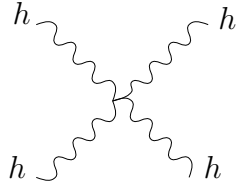}%
\end{picture}%
\setlength{\unitlength}{0.00033300in}%
\begingroup\makeatletter\ifx\SetFigFont\undefined
\def\x#1#2#3#4#5#6#7\relax{\def\x{#1#2#3#4#5#6}}%
\expandafter\x\fmtname xxxxxx\relax \def\y{splain}%
\ifx\x\y   
\gdef\SetFigFont#1#2#3{%
  \ifnum #1<17\tiny\else \ifnum #1<20\small\else
  \ifnum #1<24\normalsize\else \ifnum #1<29\large\else
  \ifnum #1<34\Large\else \ifnum #1<41\LARGE\else
     \huge\fi\fi\fi\fi\fi\fi
  \csname #3\endcsname}%
\else
\gdef\SetFigFont#1#2#3{\begingroup
  \count@#1\relax \ifnum 25<\count@\count@25\fi
  \def\x{\endgroup\@setsize\SetFigFont{#2pt}}%
  \expandafter\x
    \csname \romannumeral\the\count@ pt\expandafter\endcsname
    \csname @\romannumeral\the\count@ pt\endcsname
  \csname #3\endcsname}%
\fi
\fi\endgroup
\begin{picture}(3662,2760)(389,-2533)
\put(3826,-2461){\makebox(0,0)[lb]{\smash{\SetFigFont{12}{14.4}{rm}$h$}}}
\put(4051,-136){\makebox(0,0)[lb]{\smash{\SetFigFont{12}{14.4}{rm}$h$}}}
\put(1051,-2461){\makebox(0,0)[rb]{\smash{\SetFigFont{12}{14.4}{rm}$h$}}}
\put(1126,-61){\makebox(0,0)[rb]{\smash{\SetFigFont{12}{14.4}{rm}$h$}}}
\end{picture}
\end{minipage}
\end{center}
\caption{Tree-level contact graph: $Q$}
\end{figure}

\begin{figure}[htb]
\begin{center}
\begin{picture}(0,0)%
\epsfig{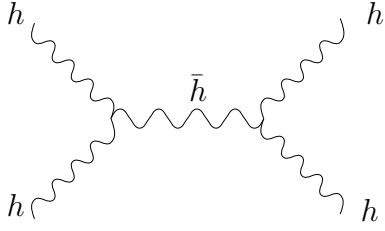}%
\end{picture}%
\setlength{\unitlength}{0.00033300in}%
\begingroup\makeatletter\ifx\SetFigFont\undefined
\def\x#1#2#3#4#5#6#7\relax{\def\x{#1#2#3#4#5#6}}%
\expandafter\x\fmtname xxxxxx\relax \def\y{splain}%
\ifx\x\y   
\gdef\SetFigFont#1#2#3{%
  \ifnum #1<17\tiny\else \ifnum #1<20\small\else
  \ifnum #1<24\normalsize\else \ifnum #1<29\large\else
  \ifnum #1<34\Large\else \ifnum #1<41\LARGE\else
     \huge\fi\fi\fi\fi\fi\fi
  \csname #3\endcsname}%
\else
\gdef\SetFigFont#1#2#3{\begingroup
  \count@#1\relax \ifnum 25<\count@\count@25\fi
  \def\x{\endgroup\@setsize\SetFigFont{#2pt}}%
  \expandafter\x
    \csname \romannumeral\the\count@ pt\expandafter\endcsname
    \csname @\romannumeral\the\count@ pt\endcsname
  \csname #3\endcsname}%
\fi
\fi\endgroup
\begin{picture}(5987,3435)(389,-3208)
\put(6376,-61){\makebox(0,0)[lb]{\smash{\SetFigFont{12}{14.4}{rm}$h$}}}
\put(6301,-3136){\makebox(0,0)[lb]{\smash{\SetFigFont{12}{14.4}{rm}$h$}}}
\put(3901,-1261){\makebox(0,0)[rb]{\smash{\SetFigFont{12}{14.4}{rm}$\bar{h}$}}}
\put(1051,-61){\makebox(0,0)[rb]{\smash{\SetFigFont{12}{14.4}{rm}$h$}}}
\put(1051,-3061){\makebox(0,0)[rb]{\smash{\SetFigFont{12}{14.4}{rm}$h$}}}
\end{picture}
\end{center}
\caption{Tree-level exchange graph: $V_{q}\Pi V_{R}$ and $V_R\tilde{\Pi}V_R$}
\end{figure}

The fourth-order contribution to the tree level contact term is given by 
\bea
Q&=&\val{\rg\{ a\;\Rie+b\;\Ric+c\; \Rs\}}{h^4}\nonumber\\
&=&s^2E_1\Big(-\frac{a}{4}+\frac{b}{8}+\frac{9}{16}c\Big)+t^2 
E_1\Big(\frac{a}{8}+\frac{b}{16}\Big)+
s^2E_2\Big(\frac{3}{2}a+\frac{b}{4}\Big)+\frac{a}{8}u^2E_2 + \cdots
\eea

The exchange graph gives rise to two terms of the order ${\cal O}(k^4)$. 
The first one contains the corrected propagator
$\tilde{\Pi}$, giving, isloating the $E_1$ and $E_2$ terms,
\be
V_R^{\mu\nu}\tilde{\Pi}_{\mu\nu\alpha\beta}V_R^{\alpha\beta}=s^2E_1
\Big(-\frac{a}{16}+\frac{b}{8}+\frac{9}{16}c\Big)+t^2E_1\,
\frac{b+4a}{16}+s^2E_2\,\frac{b+4a}{4} + \cdots
\ee
and the second exchange diagram has the usual graviton propagator as the 
internal 
propagator
\be
V_q^{\mu\nu}\Pi_{\mu\nu\alpha\beta}V_R^{\alpha\beta}=s^2E_1\Big(\frac{a}{4}
-\frac{b}{4}-\frac{9}{8}c\Big)+t^2E_1\Big(
-\frac{a}{4}-\frac{b}{8}\Big)+s^2E_2\Big(-\frac{5}{2}a-\frac{b}{2}\Big) 
+ \cdots
\ee
where $V_q^{\mu\nu}=a{\VRie}+b{\VRic}+c{\VRs}$ is 
the vertex for the quadratic curvature
terms.
If we add up everything, retaining only the $E_1$ and $E_2$ terms, 
we get the following result, 
\be
Q+V_R^{\mu\nu}\tilde{\Pi}_{\mu\nu\alpha\beta}V_R^{\alpha\beta}
+V_q^{\mu\nu}\Pi_{\mu\nu\alpha\beta}V_R^{\alpha\beta}=
\frac{a}{8}\Big\{\Big(-\frac{1}{2}s^2+t^2\Big)E_1+u^2 E_2\Big\} =
\frac{a}{8} K_{4g}^{tree}
\ee
This result tells us  that  tree-level string amplitudes are reproduced by the 
Riemann squared term
only and  that the coefficients $b$ and  $c$, which are associated with the 
Ricci squared and the scalar curvature squared term 
respectively, do not appear. They cancel during the summation
over contact and exchange 
diagrams. It follows that the on-shell string amplitudes do not
fix a particular combination of curvature squared terms. 

Before discussing the consequences of this result, let us consider the 
situation for the genus-one case.
Similar to tree level, there is also one contact term, Fig.3, which gives 
$(\Delta_{gr} Q)$, where 
\be
\Delta_{gr}=\frac{1}{i\pi}\sum_{{\bf s}\,even} 
(-)^{s_1+s_2}\int\frac{d^2\tau}{\tau_2}\frac{1}{|\eta(\tau)|^4}\,Z_{int}\,
\p_{\tau}Z_{\psi}\,\hat{G}_2
(\bar{\tau})
\ee
is replacing the blob representing one-loop 1PI processes.
\begin{figure}[htb]
\begin{center}
\begin{picture}(0,0)%
\epsfig{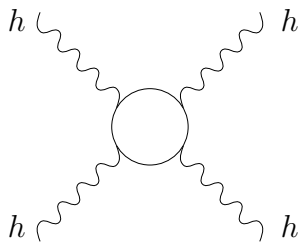}%
\end{picture}%
\setlength{\unitlength}{0.00033300in}%
\begingroup\makeatletter\ifx\SetFigFont\undefined
\def\x#1#2#3#4#5#6#7\relax{\def\x{#1#2#3#4#5#6}}%
\expandafter\x\fmtname xxxxxx\relax \def\y{splain}%
\ifx\x\y   
\gdef\SetFigFont#1#2#3{%
  \ifnum #1<17\tiny\else \ifnum #1<20\small\else
  \ifnum #1<24\normalsize\else \ifnum #1<29\large\else
  \ifnum #1<34\Large\else \ifnum #1<41\LARGE\else
     \huge\fi\fi\fi\fi\fi\fi
  \csname #3\endcsname}%
\else
\gdef\SetFigFont#1#2#3{\begingroup
  \count@#1\relax \ifnum 25<\count@\count@25\fi
  \def\x{\endgroup\@setsize\SetFigFont{#2pt}}%
  \expandafter\x
    \csname \romannumeral\the\count@ pt\expandafter\endcsname
    \csname @\romannumeral\the\count@ pt\endcsname
  \csname #3\endcsname}%
\fi
\fi\endgroup
\begin{picture}(4637,3616)(89,-3069)
\put(4726,-2986){\makebox(0,0)[lb]{\smash{\SetFigFont{12}{14.4}{rm}$h$}}}
\put(4726,239){\makebox(0,0)[lb]{\smash{\SetFigFont{12}{14.4}{rm}$h$}}}
\put(751,-2986){\makebox(0,0)[rb]{\smash{\SetFigFont{12}{14.4}{rm}$h$}}}
\put(751,239){\makebox(0,0)[rb]{\smash{\SetFigFont{12}{14.4}{rm}$h$}}}
\end{picture}
\end{center}
\caption{1-loop-level contact graph: $(\Delta_{gr} Q)$}
\end{figure}

There is no wavefunction renormalization of on-shell external legs, 
because we have $\val{\rg R}{h\bh}=0$,
$\val{\rg\Rie}{h\bh}=0$ and the same is true for the Ricci squared and the 
scalar curvature squared term. Hence, Fig.4 does
not contribute to the curvature squared term in the effective action.
\begin{figure}[htb]
\begin{center}
\begin{picture}(0,0)%
\epsfig{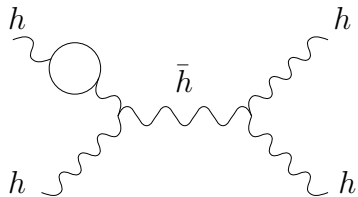}%
\end{picture}%
\setlength{\unitlength}{0.00033300in}%
\begingroup\makeatletter\ifx\SetFigFont\undefined
\def\x#1#2#3#4#5#6#7\relax{\def\x{#1#2#3#4#5#6}}%
\expandafter\x\fmtname xxxxxx\relax \def\y{splain}%
\ifx\x\y   
\gdef\SetFigFont#1#2#3{%
  \ifnum #1<17\tiny\else \ifnum #1<20\small\else
  \ifnum #1<24\normalsize\else \ifnum #1<29\large\else
  \ifnum #1<34\Large\else \ifnum #1<41\LARGE\else
     \huge\fi\fi\fi\fi\fi\fi
  \csname #3\endcsname}%
\else
\gdef\SetFigFont#1#2#3{\begingroup
  \count@#1\relax \ifnum 25<\count@\count@25\fi
  \def\x{\endgroup\@setsize\SetFigFont{#2pt}}%
  \expandafter\x
    \csname \romannumeral\the\count@ pt\expandafter\endcsname
    \csname @\romannumeral\the\count@ pt\endcsname
  \csname #3\endcsname}%
\fi
\fi\endgroup
\begin{picture}(5537,2925)(314,-2536)
\put(5851,-2461){\makebox(0,0)[lb]{\smash{\SetFigFont{12}{14.4}{rm}$h$}}}
\put(5776, 89){\makebox(0,0)[lb]{\smash{\SetFigFont{12}{14.4}{rm}$h$}}}
\put(3301,-886){\makebox(0,0)[lb]{\smash{\SetFigFont{12}{14.4}{rm}$\bar{h}$}}}
\put(976, 89){\makebox(0,0)[rb]{\smash{\SetFigFont{12}{14.4}{rm}$h$}}}
\put(976,-2461){\makebox(0,0)[rb]{\smash{\SetFigFont{12}{14.4}{rm}$h$}}}
\end{picture}
\end{center}
\caption{Wavefunction renormalization of external legs}
\end{figure}

Furthermore, we have one contribution from Fig.5 which is given by
$(\Delta_{gr}\tilde{V}_q^{\mu\nu})\Pi_{\mu\nu\alpha\beta}\VR^{\alpha\beta}$,
where the factor $\frac{1}{2}$ reflects the fact that there is no 
renormalization of Newton's constant
and $\tilde{V}_q^{\mu\nu}$ stands for the one-loop curvature squared term 
with one-loop 
coefficients $a$, $b$ and $c$.

\begin{figure}[htb]
\begin{center}
\begin{picture}(0,0)%
\epsfig{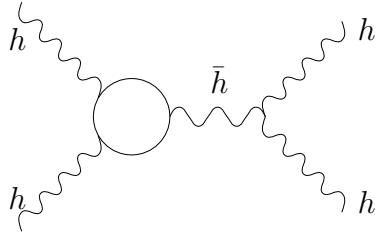}%
\end{picture}%
\setlength{\unitlength}{0.00033300in}%
\begingroup\makeatletter\ifx\SetFigFont\undefined
\def\x#1#2#3#4#5#6#7\relax{\def\x{#1#2#3#4#5#6}}%
\expandafter\x\fmtname xxxxxx\relax \def\y{splain}%
\ifx\x\y   
\gdef\SetFigFont#1#2#3{%
  \ifnum #1<17\tiny\else \ifnum #1<20\small\else
  \ifnum #1<24\normalsize\else \ifnum #1<29\large\else
  \ifnum #1<34\Large\else \ifnum #1<41\LARGE\else
     \huge\fi\fi\fi\fi\fi\fi
  \csname #3\endcsname}%
\else
\gdef\SetFigFont#1#2#3{\begingroup
  \count@#1\relax \ifnum 25<\count@\count@25\fi
  \def\x{\endgroup\@setsize\SetFigFont{#2pt}}%
  \expandafter\x
    \csname \romannumeral\the\count@ pt\expandafter\endcsname
    \csname @\romannumeral\the\count@ pt\endcsname
  \csname #3\endcsname}%
\fi
\fi\endgroup
\begin{picture}(5837,3616)(89,-2769)
\put(5926,-2461){\makebox(0,0)[lb]{\smash{\SetFigFont{12}{14.4}{rm}$h$}}}
\put(5926,239){\makebox(0,0)[lb]{\smash{\SetFigFont{12}{14.4}{rm}$h$}}}
\put(3901,-586){\makebox(0,0)[rb]{\smash{\SetFigFont{12}{14.4}{rm}$\bar{h}$}}}
\put(751, 89){\makebox(0,0)[rb]{\smash{\SetFigFont{12}{14.4}{rm}$h$}}}
\put(751,-2386){\makebox(0,0)[rb]{\smash{\SetFigFont{12}{14.4}{rm}$h$}}}
\end{picture}
\end{center}
\caption{Contribution from: $(\Delta_{gr}\,\tilde{V}_q)\Pi V_R$}
\end{figure}

Finally there are three terms of the order ${\cal O}(k^4)$ coming 
from a diagram with 
an internal blob like Fig.6. 
\begin{figure}[htb]
\begin{center}
\begin{picture}(0,0)%
\epsfig{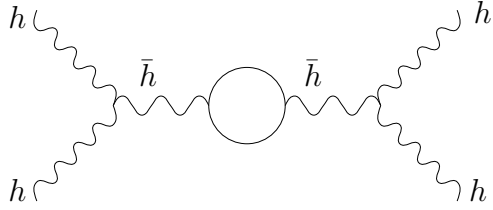}%
\end{picture}%
\setlength{\unitlength}{0.00033300in}%
\begingroup\makeatletter\ifx\SetFigFont\undefined
\def\x#1#2#3#4#5#6#7\relax{\def\x{#1#2#3#4#5#6}}%
\expandafter\x\fmtname xxxxxx\relax \def\y{splain}%
\ifx\x\y   
\gdef\SetFigFont#1#2#3{%
  \ifnum #1<17\tiny\else \ifnum #1<20\small\else
  \ifnum #1<24\normalsize\else \ifnum #1<29\large\else
  \ifnum #1<34\Large\else \ifnum #1<41\LARGE\else
     \huge\fi\fi\fi\fi\fi\fi
  \csname #3\endcsname}%
\else
\gdef\SetFigFont#1#2#3{\begingroup
  \count@#1\relax \ifnum 25<\count@\count@25\fi
  \def\x{\endgroup\@setsize\SetFigFont{#2pt}}%
  \expandafter\x
    \csname \romannumeral\the\count@ pt\expandafter\endcsname
    \csname @\romannumeral\the\count@ pt\endcsname
  \csname #3\endcsname}%
\fi
\fi\endgroup
\begin{picture}(7637,3150)(58,-2536)
\put(7695,314){\makebox(0,0)[lb]{\smash{\SetFigFont{12}{14.4}{rm}$h$}}}
\put(7620,-2461){\makebox(0,0)[lb]{\smash{\SetFigFont{12}{14.4}{rm}$h$}}}
\put(2476,-661){\makebox(0,0)[lb]{\smash{\SetFigFont{12}{14.4}{rm}$\bar{h}$}}}
\put(5326,-661){\makebox(0,0)[rb]{\smash{\SetFigFont{12}{14.4}{rm}$\bar{h}$}}}
\put(720,-2461){\makebox(0,0)[rb]{\smash{\SetFigFont{12}{14.4}{rm}$h$}}}
\put(720,239){\makebox(0,0)[rb]{\smash{\SetFigFont{12}{14.4}{rm}$h$}}}
\end{picture}
\end{center}
\caption{Contributions from: $V_q\Pi(\Delta_{gr} \bar{V}_R)\Pi V_R$ and $V_R\Pi 
(\Delta_{gr}\bar{V}_R)
\tilde{\Pi}V_R$ and $V_R\Pi (\Delta_{gr}\bar{V}_q)\Pi V_R$}
\end{figure}
Before giving
the explicit expression of these terms, we compute the kinematic expression 
for the field theory two-point function with two external
off shell legs $\bh$ for various curvature terms. The expansion of $\val{\rg 
R}{\bh^2}$ gives 
$(\bar{V}_R)_{\mu\nu\rho\sigma}=\frac{s}{4}\Pi_{\mu\nu\rho\sigma}$ and 
$\val{\rg\{a\Rie+b\Ric+c\Rs\}}{\bh^2}$ leads to 
$(\bar{V}_q)_{\mu\nu\rho\sigma}=\frac{s^2}{4}\tilde{\Pi}_{\mu\nu\rho\sigma}$. 
The term  
\be
V_q^{\alpha\beta}\Pi_{\alpha\beta\mu\nu}(\Delta_{gr}\bar{V}_R^{\mu\nu\gamma\del}
)\Pi_{\gamma\del\rho\sigma}V_R^{\rho\sigma}
\ee
 does not give any 
contribution to the one-loop effective action
because of the tree level vertex $V_q^{\alpha\beta}$ depends on tree-level 
coefficients.
There are two more ${\cal O}(k^4)$ contributions, namely 
\bea
\VR^{\alpha\beta}\Pi_{\alpha\beta\mu\nu}(\Delta_{gr}
\bar{V}_R^{\mu\nu\gamma\delta})\tilde{\Pi}_{\gamma\delta\rho\sigma}\VR^
{\rho\sigma}&=& -\frac{\Delta_{gr}}{4}\Big\{(b+4 a) 
(\VR)_{\rho\sigma}\VR^{\rho\sigma}-\frac{(b+5 a-c)}{4} \VR^2\Big\}\\
\VR^{\alpha\beta}\Pi_{\alpha\beta\mu\nu}(\Delta_{gr}
\bar{V}_q^{\mu\nu\gamma\delta})\Pi_{\gamma\delta\rho\sigma}
\VR^{\rho\sigma}&=&\frac{\Delta_{gr}}{4}\Big\{(b+4 a) (\VR)_{\rho\sigma}\VR^
{\rho\sigma}-\frac{(b+5 a-c)}{4}\VR^2\Big\}
\eea
 These two exchange graphs  cancel against each other.
Adding up the exchange graphs with one blob at the three point vertex and the 
contact term, we get 
\bea
(\Delta_{gr}Q)+ 
\frac{1}{2}(\Delta_{gr}\tilde{V}_q^{\alpha\beta})\Pi_{\alpha\beta\rho\sigma}
\VR^{\rho\sigma}
&= & \Delta_{gr}\Big\{\frac{a}{8} (-s^2 E_1+(2 s^2+u^2) E_2)\Big\}\\
&=& \Delta_{gr}\frac{a}{8} K_{4g}^{1-loop}
\eea
The only one-loop coefficient that survives the field theory subtractions is,
as in the tree-level case, $a$, 
the coefficient of $\rg\Rie$. All other coefficients
cancel against each other.
Therefore, at the genus-one level, as in the genus-zero case, 
the coefficients $b$ and $c$ are not determined. They remain completely
ambiguous. Thus, for example, the topological 
Gau{\ss}-Bonnet ($GB$) term, which has $b=-4a$ and $c=a$,
and the $\C$ term, with $b=-2a$ and $c=-\frac{1}{3}$, 
produce the same kinematic expression in the string $S$ matrix.

Thus we learn that on-shell string amplitudes can only fix the coefficient
of $\rg \Rie$ as an unambiguous coefficient, whereas the coefficients of 
$\rg \Ric$ and $\rg \Rs$ are
completely ambiguous. They cancel when considering both contact and exchange 
graphs and the situation for the genus-$1$ case is exactly the same as 
for the genus-$0$ case. We expect this situation to continue at any higher
genus as well. We conclude that one could use any one of a continuously infinite
set of curvature squared terms, indexed by arbitrary coefficients $b$ and $c$,
in the string effective action, because the string 
$S$ matrix would see no
difference between them. This means that field redefinitions are a 
symmetry of the perturbative string $S$ matrix. 
This is precisely the content of the equivalence theorem \cite{iz}, \cite{ts}, 
\cite{met}, \cite{haag}, which claims that, for example, redefining the metric
$\tilde{g}_{\mu\nu}=g_{\mu\nu}+c_1 \,R_{\mu\nu}+c_2 \,g_{\mu\nu} R$ will change
the coefficients $b$ and $c$ of the effective Lagrangian ${\cal L}$ so that 
$\delta 
a=0$, $\delta b=c_1$ and $\delta c=-\frac{c_1}{2}-c_2$, 
but does not change the $S$ matrix; that is $S=\tilde{S}$. Our results
explicitly verify previous results on field redefinitions. We also
show exactly how this ambiguity arises within individual genus-zero and
genus-one amplitudes.

It is important to emphasize that an effective action with only a $GB$ term, 
for example, is physically different from an effective action involving 
$R^{2}$ or $\C$. In the pure $GB$ case, the theory represents only a spin-2 
graviton whereas the $R^{2}$ case has , in addition to the spin-2 graviton, an
additional ghost-free scalar degree of freedom. Similarly, a $\C$ term in
the effective action would add a symmetric tensor of ghost-like fields to the
spectrum. These effective actions are physically inequivalent, and it is not
possible to change one into the other by a field redefinition or any other
means. It is therefore of importance to know the complete higher-derivative
structure of the superstring effective action. Unfortunately, the results of
this paper show that it is not possible to determine this structure by
computing on-shell, $S$-matrix string amplitudes. Such computations can always
be reproduced by an infinite number of effective actions indexed by
coefficients $b$ and $c$, and this ambiguity can never be resolved in this
manner. This does not mean, however, that one can use field redefinitions to
choose some convenient higher-derivative action, such as pure $GB$. It simply
means that one would have to perform some kind of off-shell superstring
calculation, such as in string field theory, to exactly determine the
structure of the higher-derivative gravitational terms.


\subsection*{4. Gravitational Threshold Corrections for Symmetric Orbifolds}

Moduli-dependent threshold corrections have been discussed in \cite{agn1}, 
\cite{ant}, \cite{kal}, by calculating a $CP$ odd amplitude.
Since we are interested in the kinematic structure of the $CP$ even part of 
the effective action corresponding to moduli
dependent gravitational corrections, 
we now want to calculate the $CP$ even string amplitude including four 
graviton vertex operators and one modulus
vertex
\be
V_{T}^{(0)}(z,\bar{z})=: v_{IJ}\,\p X^I (\p X^J+i k\cdot\psi\psi^J)\,e^{ik\cdot 
X}:
\ee
where $v_{IJ}=\p_T(G_{IJ}+B_{IJ})$, $G_{IJ}$ is the metric and $B_{IJ}$ the 
antisymmetric tensor of the six dimensional 
internal compact manifold. The five-point amplitude is
\be
\la\Pi_{i=1}^4V_{g}^{(0)}(z_i,\bar{z}_i)\,V_{T}^{(0)}(z_5,\bar{z}_5)\ra
=v_{IJ} \la\bar{\p}X^I\p X^J\ra\la\Pi_{i=1}^4V_{g}^{(0)}(z_i,
\bar{z}_i)\ra
\ee
which can be computed to give the result
\be
{\cal A}_{4g1T}^{even}=\frac{6 g_s^5}{16^4} v_{IJ}\,K_{4g}^{1-loop}\int 
d^2\tau\,{\cal B}^{IJ}_{gr}(\tau,\bar{\tau})
\ee
where 
\be
{\cal B}_{gr}^{IJ}(\tau,\bar{\tau})=\frac{1}{i\pi}\la\bar{\p}X^I\,\p 
X^J\ra\,\sum_{{\bf s}\,even}(-)^{s_1+s_2}\frac{1}{|\eta(\tau)|^4}
\,Z_{int}\,\p_{\tau}Z_{\psi}\,\hat{G}_2(\bar{\tau}).
\ee
It is only the $N=2$ sector that gives a non-vanishing contribution to ${\cal 
B}_{gr}^{IJ}$, because in this sector the zero
modes of $X^I$ can be saturated.

If we write ${\cal B}_{gr}^{IJ}(\tau,\bar{\tau})=\la\bar{\p}X^I\p X^J\ra\,
{\cal B}_{gr}(\tau,\bar{\tau})$ we have a similar expression
to the threshold corrections to gauge couplings \cite{kal}. Therefore, we can 
express ${\cal B}_{gr}$ in factorized form as
${\cal B}_{gr}=Z_{torus}(\tau,\bar{\tau})\,{\cal C}_{gr}(\bar{\tau})$, where 
$Z_{torus}=\sum_{p_L,p_R\in\Gam_{2,2}}q^{p_L^2/2}
\bar{q}^{p_R^2/2}$ is the partition function of the zero modes of $X^I$ and 
$\Gam_{2,2}$ is an even self dual Lorentzian lattice.
The coupling appearing in the five-point amplitude is thus 
\be
\p_{T}\Delta_{gr}=\frac{1}{i\pi}\int\frac{d^2\tau}{2\pi\tau_2}\p_TZ_{torus}\,
{\cal C}_{gr}(\bar{\tau})
\ee
If we apply $\p_{\bar{T}}$ on both sides of this equation and use the fact 
that 
\be
\p_{\bar{T}}\p_T Z_{torus}=\frac{4\tau_2}{(T+\bar{T})^2}
\p_{\bar{\tau}}\p_{\tau}(\tau_2 Z_{torus})
\ee
 \cite{agn1}, we can perform the 
$\tau$ integral which becomes a contour integral
around the fundamental region.The only non-vanishing contribution comes
from the $\tau_2\to\infty$ region, because
of the modular invariance of the integrand. Thus we get directly the result
\be
\p_{\bar{T}}\p_T\Delta_{gr}=\frac{b_{gr}}{\pi^2(T+\bar{T})^2}
\ee
where $b_{gr}=\lim_{\tau_2\to\infty}{\cal B}_{gr}$ is the trace anomaly for 
the $N=2$ sector. Integrating the last expression
yields to the gravitational coupling of the curvature squared terms
\begin{equation} 
\Delta_{gr}(T,\bar{T})=-\frac{b_{gr}}{\pi^2}\ln\Big(|\eta(iT)|^4(T+\bar{T})
\Big)
\label{eq:hello}
\end{equation} 

If we now use the results of the last section,   we know that the moduli 
dependent 
threshold correction $\Delta_{gr}$ is the coefficient
of $\rg \Rie$. There could also be $\Ric$ or $\Rs$ terms
whose coefficients
are ambiguous and cannot be determined by string amplitude calculations.
Therefore, this part of the one-loop 
string effective Lagrangian is given by
\be
{\cal L}=\rg \Delta_{gr} (\Rie + b\;\Ric +c\; \Rs)
\label{eq:hi}
\ee
As discussed above, it would require some sort of off-shell superstring
calculation to fix the $b$ and $c$ coefficients uniquely. Expression 
(\ref{eq:hi}) is the best that one can compute using on-shell string
amplitudes.

\vspace{2cm}
{\bf Acknowledgement}: K.F would like to thank J. Louis and V. Kaplunovsky for 
useful discussions and the University of Pennsylvania for hospitality.
This work was supported in part by DOE Contract DOE-AC02-76-ERO-3071 and 
NATO Grant CRG. 940784 and GIF-the German-Israeli Foundation.


\end{document}